\documentclass[aip,apl,reprint]{revtex4-1}
\usepackage{graphicx}
\usepackage{dcolumn}
\usepackage{bm}
\usepackage[utf8]{inputenc}
\usepackage[T1]{fontenc}
\usepackage{mathptmx}
\usepackage[dvipsnames]{xcolor}
\usepackage[normalem]{ulem}
\begin{document}
\bibliographystyle{apsrev4-1}

\title[Fluorescence Detection of a Trapped Ion with a Monolithically Integrated Single-Photon-Counting Avalanche Diode]{Fluorescence Detection of a Trapped Ion with a Monolithically Integrated Single-Photon-Counting Avalanche Diode}

\author{W.~J.~Setzer}
\email{wjsetze@sandia.gov}
\affiliation{Sandia National Laboratories, Albuquerque, New Mexico 87185, USA}
\author{M.~Ivory}
\affiliation{Sandia National Laboratories, Albuquerque, New Mexico 87185, USA}
\author{O.~Slobodyan}
\affiliation{Sandia National Laboratories, Albuquerque, New Mexico 87185, USA}
\author{J.~W.~Van~Der~Wall}
\affiliation{Sandia National Laboratories, Albuquerque, New Mexico 87185, USA}
\author{L.~P.~Parazzoli}
\affiliation{Sandia National Laboratories, Albuquerque, New Mexico 87185, USA} 
\author{D.~Stick}
\affiliation{Sandia National Laboratories, Albuquerque, New Mexico 87185, USA} 
\author{M.~Gehl}
\affiliation{Sandia National Laboratories, Albuquerque, New Mexico 87185, USA} 
\author{M.~G.~Blain}
\affiliation{Sandia National Laboratories, Albuquerque, New Mexico 87185, USA} 
\author{R.~R.~Kay}
\affiliation{Sandia National Laboratories, Albuquerque, New Mexico 87185, USA}
\author{H.~J.~McGuinness}
\affiliation{Sandia National Laboratories, Albuquerque, New Mexico 87185, USA} 
 
\date{\today}

\begin{abstract}
We report on the demonstration of fluorescence detection from a trapped ion using single-photon avalanche photodiodes (SPADs) monolithically integrated with a microfabricated surface ion trap. The SPADs are located below the trapping positions of the ions and designed to detect 370~nm photons emitted from single $^{174}$Yb$^+$ and $^{171}$Yb$^+$ ions. We achieve an ion/no-ion detection fidelity for $^{174}$Yb$^+$ of 0.99 with an average detection window of 7.7(1)~ms. We report a dark count rate as low as 1.2~kcps for room temperature operation. The fidelity is limited by laser scatter, dark counts, and heating that prevents holding the ion directly above the SPAD. We measure count rates from each of the contributing sources and fluorescence as a function of ion position. We use the ion as a calibrated light source along with measurements of the active detector area to estimate a SPAD quantum efficiency of 24$\pm$1\%. 
\end{abstract}
\maketitle
\par Quantum technologies are rapidly advancing with promising applications such as computing and simulation, networking, sensing, and precision timekeeping. One of the primary challenges for future quantum devices is the ability to scale up the number of qubits. For ion trap systems, scaling requires the ability to independently optically address and detect many qubits. Recent work has shown scalable optical addressing based on trap-integrated photonic waveguides \cite{Mehta2016,Niffenegger2020,Mehta2020,Ivory2020}, and in this paper we demonstrate a trap-integrated scalable technology for qubit detection. 
\par Typically, a trapped ion’s quantum state is read out by collecting the fluorescence from a laser-driven optical cycling transition with a lens that directs photons to a detector. Photon collection efficiency (the fraction of emitted light that hits the detector) is often low, usually $\sim$1-2\%. More light can be collected using higher numerical aperture (NA) lenses, however these optics are large and may require short working distances, which interfere with ion trapping. Imaging systems with lenses and detectors external to the vacuum chamber have achieved collection efficiencies as high as 10\% for optics with NA~$\simeq$~0.6 \cite{Myerson2008, Burrell2010, Christensen2020, Zhukas2020, Noek2013} and readout fidelities above 0.999 and approaching 0.9999.
\par Commonly used external detectors for trapped-ion systems, such as photomultiplier tubes (PMT), electron multiplying charge coupled device (EMCCD) cameras, and avalanche photodiodes (APD), have quantum efficiencies of 20-40\% for blue to UV photons. Superconducting nanowire single photon detectors (SNSPDs), which require cryogenic temperatures for operation, have achieved impressive efficiencies as high as 80\%\cite{Slichter2017, Wollman2017} for UV photons and measured the state of a $^{171}$Yb$^+$ qubit with fidelity >0.999 in 11~$\mu$s \cite{Crain2019}. Considering both the collection efficiency and quantum efficiency of these external detectors, the best total detection efficiencies demonstrated range from 2-4\%\cite{Crain2019,Noek2013}. 
\par These traditional imaging systems may support scaling for modular trapped-ion architectures \cite{Brown2016}, but they are not scalable for imaging arbitrarily extended arrays of ions \cite{Kielpinski2002} because the lateral dimensions of the lens exceeds the lateral dimensions of the ion array it images. A promising solution to the challenge of achieving both scalability and high collection efficiency is on-chip state readout \cite{Bruzewicz2019}. Fluorescence detection has been demonstrated using several near-ion fluorescence collection techniques with components internal to the vacuum chamber: optical fibers\cite{Clark2014,Takahashi2013,VanDevender2010}, reflective traps\cite{Herskind2011,VanRynbach2016}, transparent traps\cite{Eltony2013}, microfabricated optics \cite{Merrill2011,Ghadimi2017,Jechow2011,Steed2011}, and optical cavities\cite{Sterk2012}. Not all of these methods are scalable, and for many of them external bulk optics and photon detectors are still required.
\par Recently, a SNSPD co-fabricated with a surface-electrode ion trap was used to detect UV light from a $^9$Be$^+$ qubit, achieving qubit state readout with 0.9991(1) fidelity in an average of 46~$\mu$s \cite{Todaro2021}. This detection scheme is scalable for large trap arrays because the entire unit is smaller than the electrodes needed to confine an ion. It was found that the SNSPD bright counts were reduced by 17\% due to the rf fields, even when the SNSPD performance was characterized using low peak amplitude rf, $\sim$8~V at 67.03~MHz.  For heavier ions like $^{171}$Yb$^+$, higher rf peak amplitudes are required for trapping and therefore present a challenge for shielding integrated SNSPD detectors from the higher amplitude rf fields. 
\par In transportable systems, such as atomic clocks, which require lower size, weight, power, and cost (SWaP-C), single-photon avalanche photodiodes (SPADs) provide an alternative type of integrated fluorescence detector \cite{Mehta2017,West2019,TICTOC} that is also compatible with CMOS fabrication. SPADs have some advantages over SNSPDs, including less sensitivity to rf voltage and low dark counts at room temperature \cite{Bronzi2012, Bronzi2013, Veerappan2016}. Here, we report on the use of a room temperature surface ion trap that is monolithically integrated with SPADs for measuring ion fluorescence.
\par A SPAD is an APD that is operated in Geiger mode, where it is biased at or beyond its breakdown voltage and is subsequently extremely sensitive to any charge entering its active area. During a detection event (i.e. a single photon impinging on the active area), the SPAD discharges its built-in over-bias potential, leading to a short dead time depending on the RC feedback of the device. Operation in Geiger mode is necessary for single photon detection, but comes with an increased dark count rate (DCR). Dark counts occur when non-photoinduced charges, spontaneously liberated by thermal effects, drift or diffuse into the active area. This results in a non-photon-induced avalanche event that is indistinguishable from a single photon detection event. 
\begin{figure}[h!]
\centering
\includegraphics[scale=0.36]{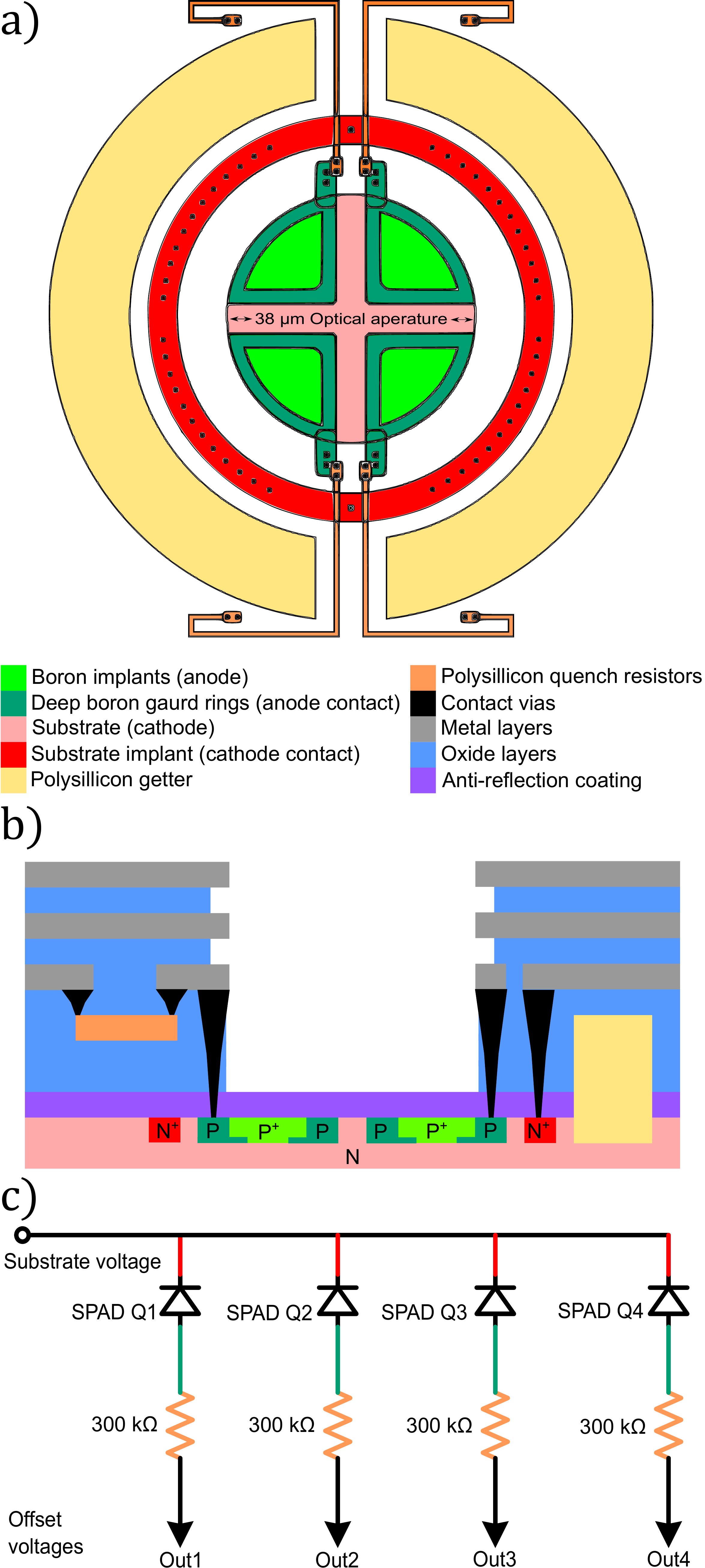}
\caption{SPAD schematic. (a) Top view showing four quartered SPADs. The SPAD anodes are recessed 7~$\mu$m below the surface of the ion trap within a 38~$\mu$m optical aperture. (b) Side view illustration. Example electrical connections between layers are shown. (c) Electrical Schematic. The quartered SPADs share the same substrate and are individually biased by varying the offset voltage of their detection circuit.} 
\label{fig:SPAD_Diagram}
\end{figure}
\par The SPADs are designed for compatibility with the geometry, material layers, and processing steps of the ion trap. The trap is fabricated monolithically after the SPAD, as shown in FIG.~\ref{fig:SPAD_Diagram}. To mitigate risk and test more devices, the circular aperture is split into quarters, each with an independent SPAD. Contacts to the anodes are made using implant tabs that are outside of the optical aperture. Silicon dioxide (SiO$_2$) thin films above the SPADs are present during fabrication and later etched away down to the anti-reflective coating (ARC). 
\par Edge breakdown in the SPADs is controlled by a deep boron diffused guard ring around the p+ active area perimeter. The deep diffusion reduces the electric field at the device periphery by increasing the radius of the p- region into the epitaxial layer. Without the edge breakdown control, the electric field lines concentrate at the abrupt edge and significantly increase dark counts. The guard ring reduces the effective active area of the device by approximately 2~$\mu$m from its edge. We determine that this unanticipated effect is due to two mechanisms. First, the deep guard ring requires a high temperature drive-in processing step that causes dopants to laterally diffuse into the region below the active area, effectively extending the width of the guard ring diffusion. Secondly, the guard ring reduces the vertical electric field near its inner edge, providing some lateral field which does not contribute to avalanche multiplication and reduces quantum efficiency.
\par While silicon is strongly absorbing at 370~nm, its large refractive index leads to significant Fresnel reflection ($\sim$57\% at normal incidence).  For this reason, we deposit an ARC over the SPADs. The ARC consists of a 10~nm silicon dioxide layer (which also functions as a Si surface passivation layer) followed by a 29~nm silicon nitride layer. This design is expected to have 10\% reflection at normal incidence.  
\par Polysilicon resistors are integrated with the SPADs to quench breakdown events and minimize dead time. The resistors are 1~$\mu$m wide and 60~$\mu$m in total length, providing about 300~k$\Omega$ per resistor. The optical aperture and anodes are surrounded by a substrate contact ring, which serves as the cathode contact to the device. Surrounding the substrate contact ring is a 10~$\mu$m wide ring of polysilicon used as an impurity getter which reduces DCR by collecting mobile impurities (such as metals) that can act as Shockley-Read-Hall generation sites, away from the SPAD junction. The ring is broken at the top and bottom to allow routing of the quench resistors. 
\par To bias the SPADs, the cathode of each SPAD is connected to a shared substrate held at a single voltage. Each SPAD's anode is connected to its own detection circuit held at a variable offset voltage. For each SPAD, the voltage difference between the offset and substrate determines its bias voltage, allowing control over which SPADs are biased above breakdown. In typical operation, the SPADs are over-biased 4~V above their 28~V breakdown voltage, with 32~V difference between their terminals. SPAD pulses are amplified and filtered by the detection circuit to remove parasitic pickup of the rf drive frequency and are converted to digital pulses by a Schmitt-trigger comparator. The resulting digital pulses are counted and timestamped with <1~ns resolution.
\begin{figure}[h!]
\centering
\includegraphics[scale=0.6]{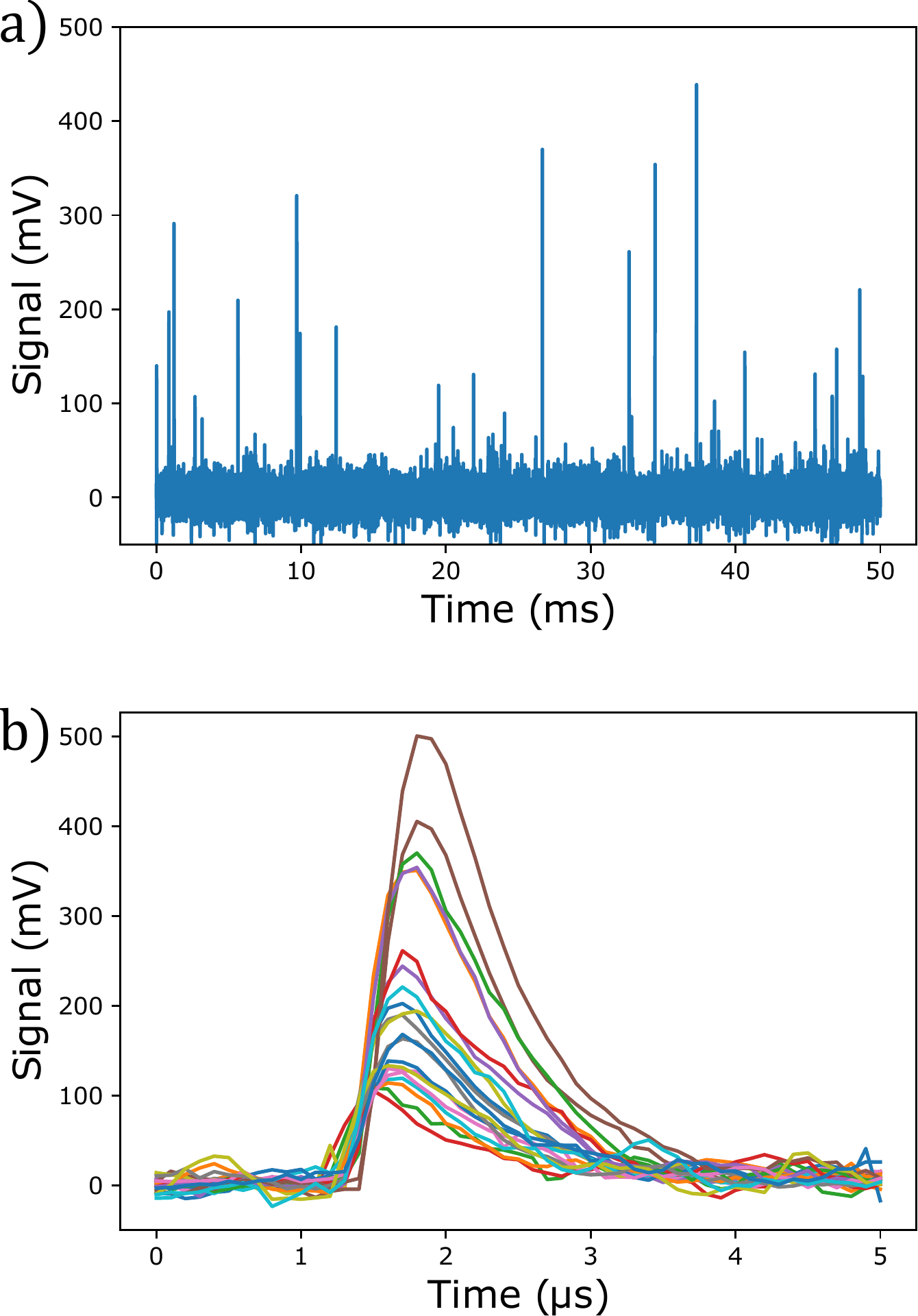}
\caption{SPAD pulses. (a) SPAD pulses probed before the Schmitt-trigger comparator in the detection circuit. (b) Stacked pulses from the 50 ms window above. Pulse amplitudes vary with overbias voltage and in typical operation range from 100--500~mV with $\sim$1~$\mu$s duration.} 
\label{fig:SPAD_Pulses}
\end{figure}
\par Oscilloscope traces acquired before digitization in the detection circuit with the trap rf and lasers off are shown in FIG.~\ref{fig:SPAD_Pulses}. Typical pulses at 32~V bias operation have 100--500~mV amplitudes and $\sim$1~$\mu$s duration. The pulse duration is not fundamentally limited by the SPAD itself and is broadened by the rf filters of the detection circuit and the capacitance of the read-out wires. A low pass filter with a 1.6 MHz cutoff is used to attenuate parasitic voltage pickup of the trap rf. Some parasitic pickup remains after filtering and contributes to counts by increasing lower voltage SPAD pulses and electrical noise above the Schmitt-trigger threshold voltage during digitization. An analysis of oscilloscope data with SPAD pulses recorded with and without the trap rf on indicates that increasing the trap rf amplitude did not contribute noticeably to additional pulses of the SPAD electrically or by increasing its temperature.
\par The dark counts of 12 individual SPADs housed within three different optical apertures on the same surface trap were measured in situ.  Dark counts vary greatly between SPADs, even those within the same optical aperture, ranging from 1.2~kcps to 140~kcps, with 6 SPAD’s having DCR below 10~kcps. Contamination during fabrication has been identified as a possible source for variation in DCR, although defects or metal contamination of the epitaxial Si cannot be ruled out.
\begin{figure}[h!]
\centering
\includegraphics[scale=0.3]{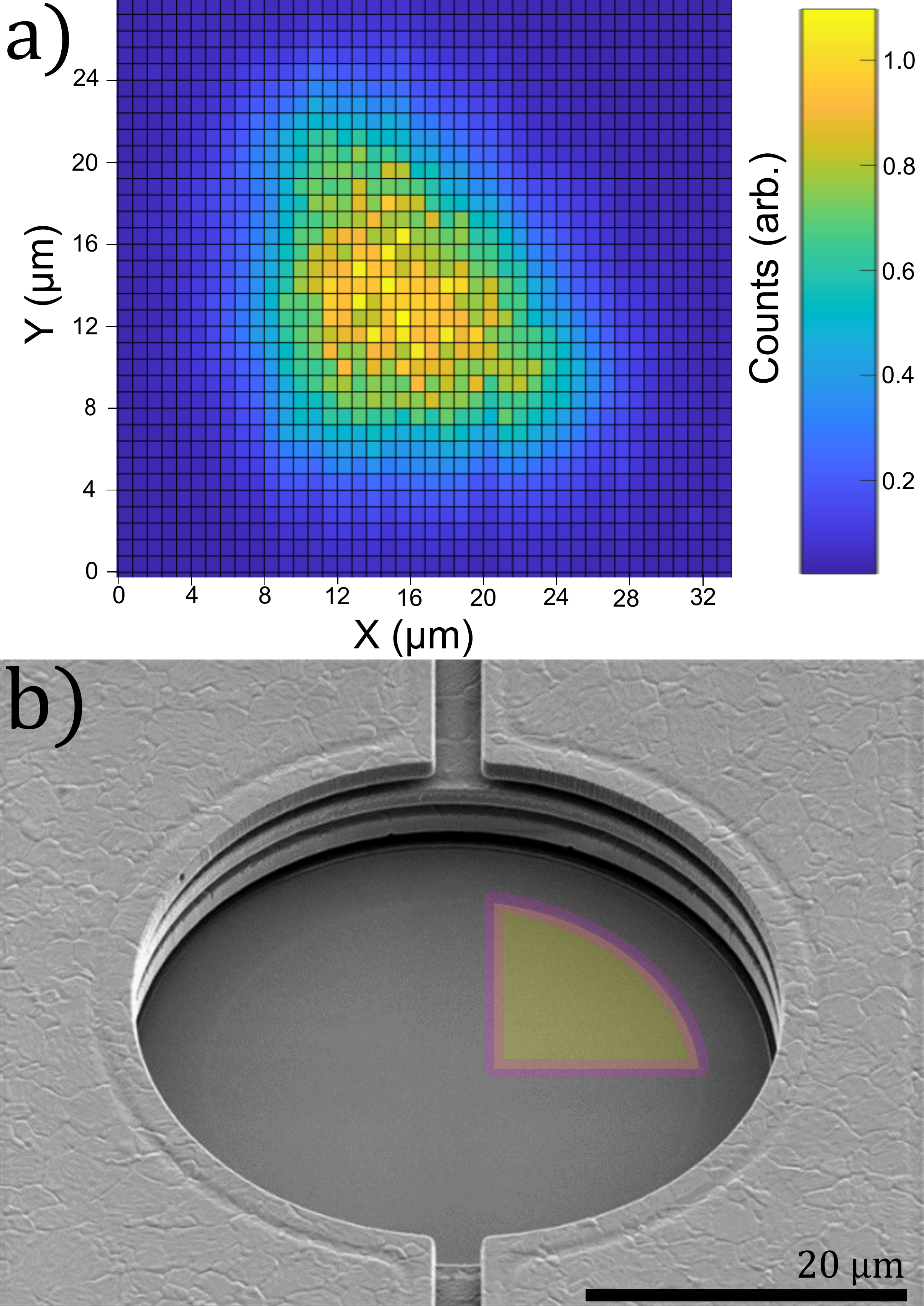}
\caption{SPAD active area. (a) Spot test measurements showing the 60~$\mu$m$^2$ active area of a single quartered SPAD, reduced due to the deep boron guard ring. (b) SEM image of the SPAD detector. False coloring indicates the active area (yellow) and the deep boron guard ring (purple) for one of the four quartered SPADs.} 
\label{fig:SPAD_Active_Area}
\end{figure}
\par We perform spot test measurements to determine the active area of the SPAD detectors. The guard ring’s effect of reducing the efficiency of bordering SPAD active area is particularly important for these quartered detectors as their area is small to begin with. These measurements are accomplished by focusing 370~nm light to a spot diameter of $\sim$1.6~$\mu$m. The spot is scanned in 800~nm steps over the region of the SPAD. For each step, the dark counts are subtracted from the total counts and the remaining counts are normalized to the maximum observed. An example of the data collected from this procedure is shown in FIG.~\ref{fig:SPAD_Active_Area}a. To ensure the optical aperture does not interfere with this tight beam, the measurement is performed with devices that have only been fabricated up through the SPAD layers.
\par The efficiency of the SPAD's active area is non-uniform due to the guard ring’s influence, as shown in FIG.~\ref{fig:SPAD_Active_Area}a. For a single quartered SPAD, we determine its effective active area to be 60~$\mu$m$^2$.  This effective active area is calculated by weighting each step's area by its normalized counts and summing those response-weighted areas. FIG.~\ref{fig:SPAD_Active_Area}b shows an SEM image of the SPADs within the optical aperture. At the edge of the optical aperture, the $\sim$1.3~$\mu$m metal layers are visible with $\sim$1~$\mu$m silicon dioxide layers hidden between them.
\par We detect 370~nm fluorescence from the $^{2}S_{1/2}~\leftrightarrow~^{2}P_{1/2}$ optical cycling transition of Yb$^+$. The $^{2}P_{1/2}$ state can decay into the long-lived $^{2}D_{3/2}$ state, which we depopulate via the $^{2}D_{3/2}~\leftrightarrow~^{3}[3/2]_{1/2}$ transition using a 935~nm repump laser. We have trapped and detected fluorescence using the SPADs from $^{171}$Yb$^+$. $^{171}$Yb$^+$ has a lower fluorescence rate due to coherent population trapping\cite{Ejtemaee2010}, therefore we use $^{174}$Yb$^+$ to characterize the SPADs. The lack of hyperfine structure in $^{174}$Yb$^+$ makes qubit state preparation and detection difficult, so we have left integrated qubit state detection for future work. 
\begin{figure}[h!]
\centering
\includegraphics[scale=0.8]{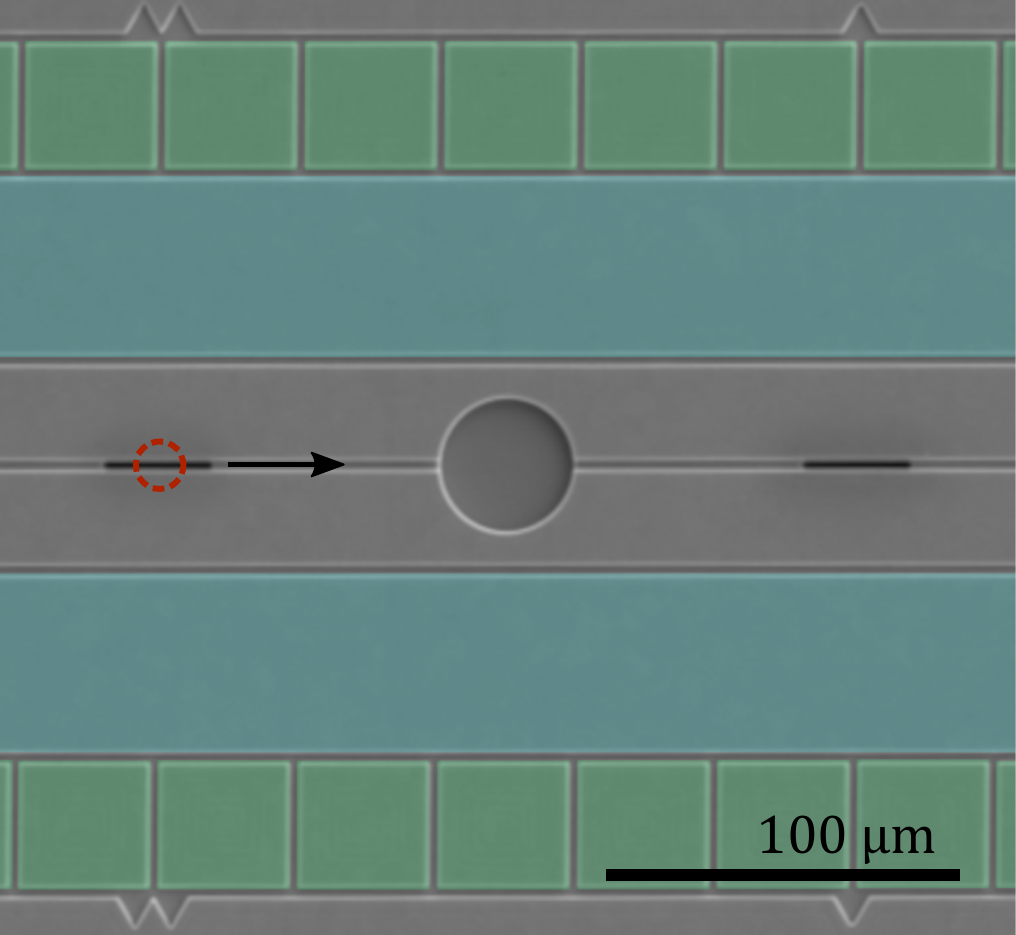}
\caption{SEM image of the ion trap with integrated SPADs. False color has been added to indicate the rf electrodes (blue) and DC electrodes (green). Ions are loaded above the slot indicated by the red circle and trapped 50~$\mu$m above the trap surface. Ions are shuttled along the trap axis towards the SPAD indicated by the black arrow.}
\label{fig:Trap}
\end{figure}
\par A false color SEM image of the surface trap with integrated SPADs used in this experiment is shown in FIG.~\ref{fig:Trap}. The trap is housed in a UHV chamber with operating pressures below 1x10$^{-11}$~Torr.  The rf electrodes, indicated in blue, provide confinement along the trap axis 50~$\mu$m above the trap surface. The DC electrodes, indicated in green, allow for ion shuttling along the rf null. Ions are loaded at loading slots and shuttled toward the detector. A hole in the trap used for loading is indicated by the red circle in FIG.~\ref{fig:Trap}. A magnetic field of 5 Gauss is applied perpendicular to the plane of the trap surface to lift the degeneracy between $^{171}$Yb$^+$ hyperfine sublevels and provide a quantization axis. This field had no discernible impact on SPAD performance. An rf drive frequency of 17.7 MHz with peak amplitude below 30~V on the rf electrodes results in typical motional frequencies of 1.3--2.0~MHz in the radial directions and an axial frequency of $\sim$450~kHz. These frequencies are verified via spectroscopy on the $|^{2}S_{1/2}~,F=0,~m_F=0\rangle$ to $|^{2}D_{3/2},~F=2,~m_F=+2\rangle$ $^{171}$Yb$^+$ quadrupole transition. 
\par The ions could not be sideband cooled, as the motional transitions decohered quickly, so heating rates were not measured. This occurred even with the SPADs unbiased and also with a gold-coated device, suggesting that this issue has other causes such as low motional frequencies or uncompensated axial micromotion.
\par We observe ion heating due to SPAD pulses when the SPADs are biased above breakdown. This is observed qualitatively by increasing the number of SPAD pulses (with laser scatter or by biasing additional SPADs above breakdown) until the ion becomes delocalized on the camera and is eventually heated out of the trap. The heating per pulse has not be quantified and therefore will need to be investigated in a future device.
\begin{table}
\caption{\label{tab:SPAD_Counts}Count Budget}
\begin{ruledtabular}
\begin{tabular}{lcr}
Source&Counts (kcps)\\
\hline
\hline
Ion Fluorescence& 4.8(1) \\
\hline
935 nm Repump Scatter& 4.0(1) \\
370 nm Doppler Scatter& 1.4(1) \\
Dark Counts& 1.2(1) \\
rf Pickup& 0.3(1)\\
\end{tabular}
\end{ruledtabular}
\end{table}
\par The SPAD counts for a typical detection experiment are shown in Table~\ref{tab:SPAD_Counts}. The count budget is acquired by toggling beams and the trap rf after the ion has been removed from the trap. In typical freespace imaging systems, an optical filter is used to block scattered 935~nm repump laser light from reaching the detector. For our integrated detector this is not possible, and so laser scatter is minimized by adjusting the position and focus of the beams. Despite minimization, scattered laser light remains the dominant source of counts outside of ion fluorescence. The laser beam intensities are adjusted to maximize fluorescence while minimizing laser scatter. For the data shown, the beam intensities are lowered until the ion fluoresces at $\sim$83\% of its saturation value. We determined this percentage by measuring and fitting the ion's fluorescence as a function of incident laser power. The rf counts result from parasitic pickup on the SPAD leads, and can be reduced in the future with improved filtering and shielding.
\begin{figure}[h!]
\centering
\includegraphics[scale=0.29]{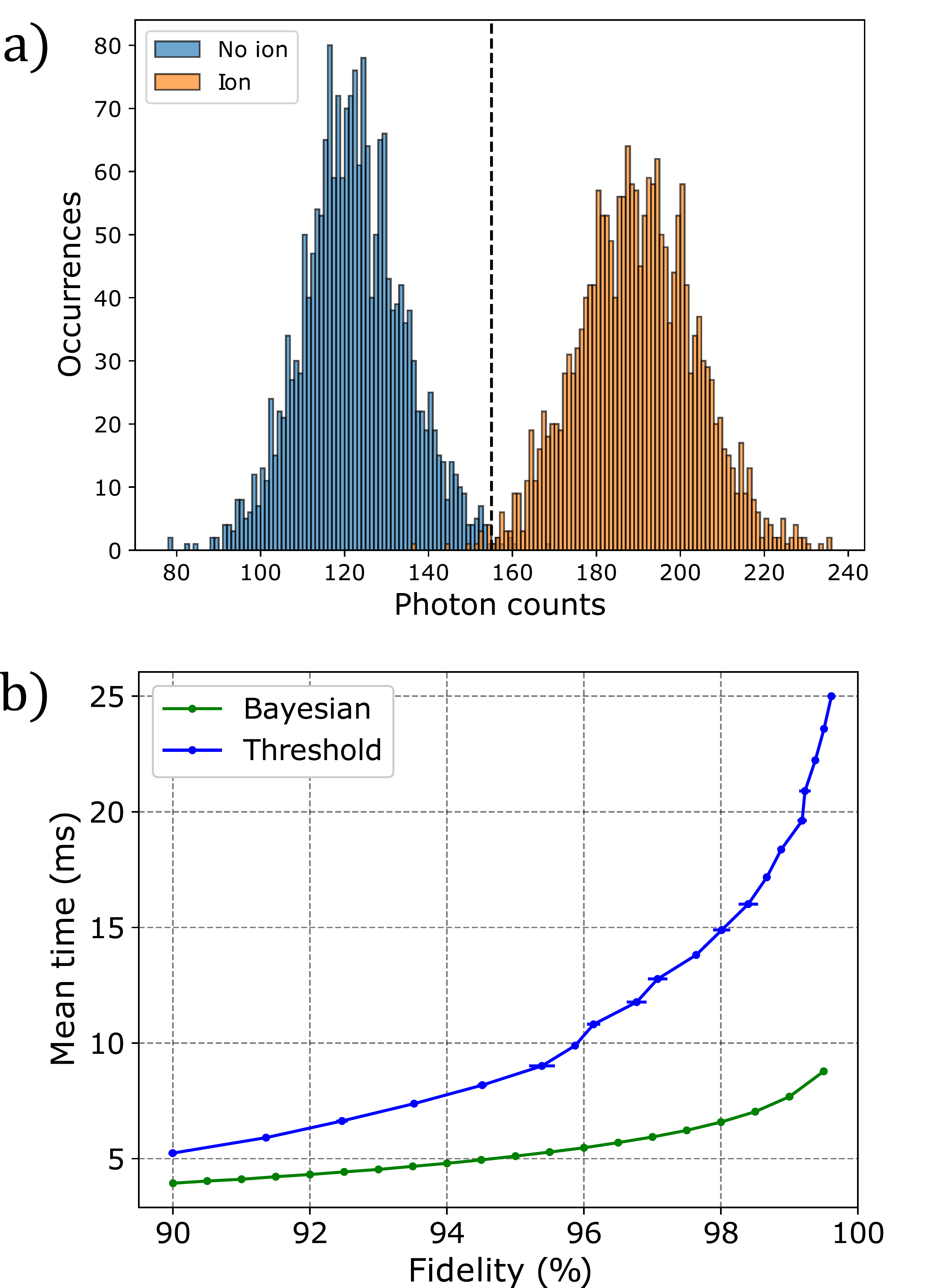}
\caption{Detection Fidelity. (a) Ion/No-ion count histogram generated using a 25~ms detection window and repeated over 50~s before (orange) and after (blue) removing the ion from the trap. The dashed line indicates the threshold for optimal fidelity of 0.996(1). (b) Fidelity plot showing average gate time to reach a target fidelity using the standard thresholding (blue) and adaptive Bayesian, technique (green). Using the adaptive Bayesian technique we measure an ion/no-ion fidelity of 0.99 with an average detection window of 7.7(1)~ms.}
\label{fig:Fidelity}
\end{figure}
\par To characterize the ion/no-ion detection fidelity, digitized pulses from a single quartered SPAD are timestamped for 50~s before and after removing the ion from the trap. The resulting timestamped datasets are then gated, counted, and binned to generate ion/no-ion count histograms. An example count histogram generated using a 25~ms gate is shown in FIG.~\ref{fig:Fidelity}a. With the timestamp datasets, we have analyzed the ion/no-ion detection fidelity using both the standard threshold technique and an adaptive Bayesian technique that allows for a variable gate time to achieve a target fidelity \cite{Myerson2008,Hume2007,Todaro2021}. We follow the technique from Myerson et al, assuming no prior information (the initial assumed probability of ion/no ion is 50\%/50\%). Because we were not performing state detection, repump dynamics did not need to be considered. Using this technique, we measure the presence of an ion with 99\% accuracy, within an average detection window of 7.7(1)~ms, as shown in FIG.~\ref{fig:Fidelity}b.
\par With the SPADs biased below breakdown, an ion can be shuttled directly above the center of the SPAD's optical aperture. However, above the breakdown voltage, SPAD pulses heat the ion out of the trap, indicating the need for additional electric field shielding, such as a mesh or an indium tin oxide (ITO) coating. This current lack of shielding limits fluorescence data collection to lateral ion positions greater than 68~$\mu$m from the center of the SPAD’s aperture. 
\begin{figure}[h!]
\centering
\includegraphics[scale=0.53]{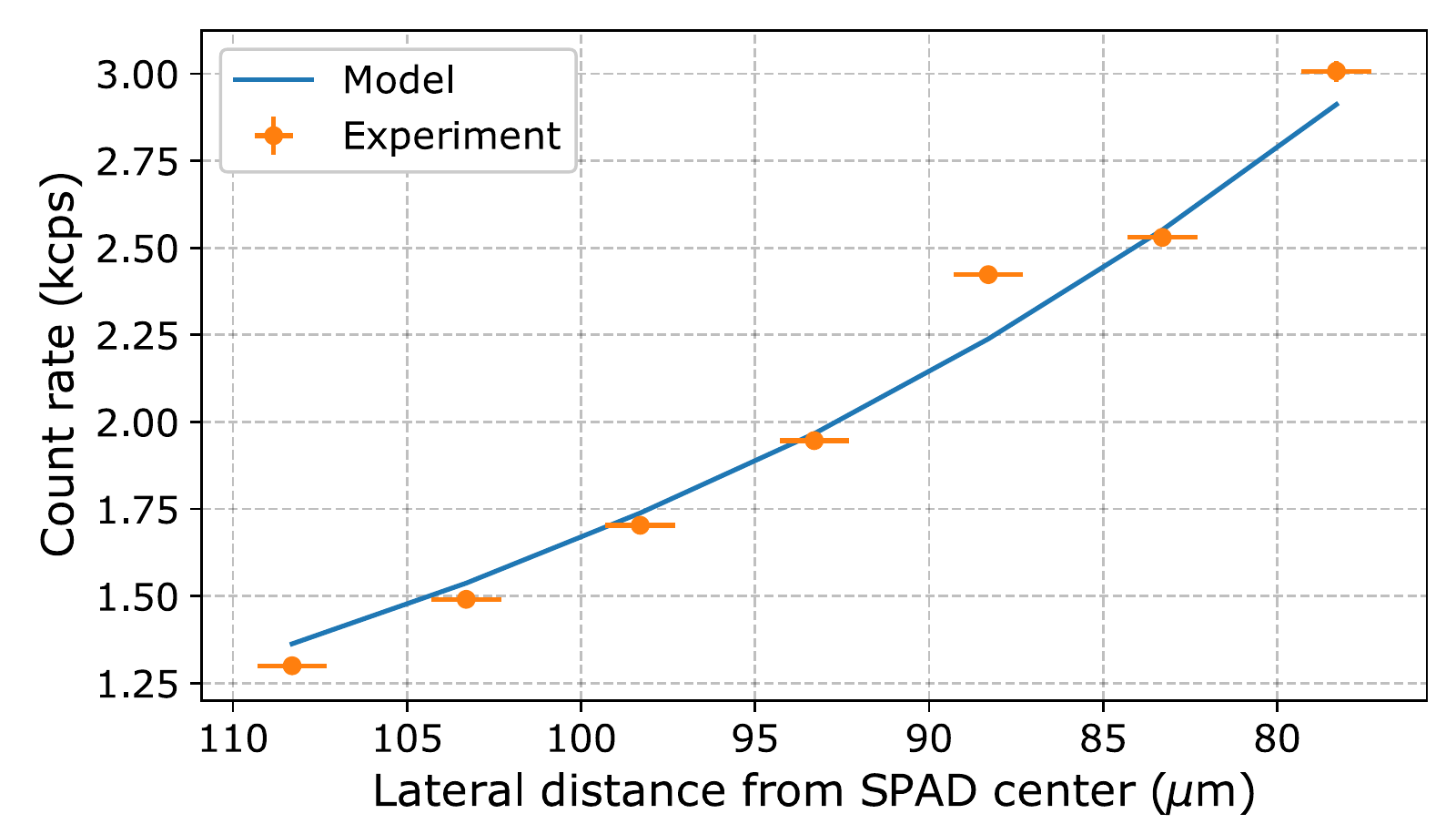}
\caption{Collection Efficiency. The fluorescence collection efficiency of the detector is modeled using the SPAD's 60~$\mu$m$^2$ effective active area (blue curve). Experimental fluorescence counts are gathered as the ion is shuttled closer to the SPAD in 5~$\mu$m lateral steps (orange points).}
\label{fig:Collection_Efficiency}
\end{figure}
\par The collection efficiency of a single quartered SPAD is modeled using the fraction of the fluorescence solid angle emission that is incident on the 60~$\mu$m$^2$ effective active area and the reflective loss of the ARC. As the angle of incidence increases, the anti-reflection minimum shifts to shorter wavelengths, increasing the amount of reflected 370 nm light. The results of this model are shown in FIG~\ref{fig:Collection_Efficiency} (blue curve). This collection efficiency is verified experimentally by collecting count data while shuttling the ion closer to the SPAD in 5~$\mu$m lateral steps along the trap axis. Fluorescence counts at each shuttling location are collected using 30 ms gating of 50~s ion/no-ion datasets (orange points in FIG~\ref{fig:Collection_Efficiency}). The reflection from the ARC ranges from 22.5\% to 17.3\% over the shuttling range shown and is as low as 10\% at normal incidence. The collection efficiency ranges with the ion's lateral distance from the SPADs center, 0.03\% at 80~$\mu$m to 0.14\% with the ion centered over the SPAD. These low collection efficiencies are primarily caused by the small active areas of the quartered SPADs and can be dramatically increased in future devices. For instance, a SPAD with an active area that fills the current optical aperture would have a collection efficiency of 2.6\%, while a trap geometry that optimizes for collection could achieve 10\% efficiency with a square SPAD that spans the distance between the rf electrodes.
\par Fluorescence is monitored during shuttling with our free-space imaging system and kept constant to $\sim$83\% of saturation by adjusting laser beam positions. The $^{174}$Yb$^+$ linewidth, $\gamma/2\pi$ = 19.6 MHz, along with the our modeled collection efficiency gives an expected incident photon rate on the active area for each ion shuttling position. The ratio of expected incident photons on the active area to the measured fluorescence counts is fit with a single parameter to measure a quantum efficiency of 24$\pm$1\%.
\par This paper describes the demonstration of ion trapping and fluorescence detection in a surface trap with monolithically integrated SPAD detectors. We measure a quantum efficiency of 24$\pm$1\% with detector dark counts as low as 1.2~kcps and detect the presence of an ion with 99\% fidelity within an average detection window of 7.7(1)~ms using an adaptive Bayesian method. This fidelity is limited primarily by detector dark counts, laser scatter, and limited ion proximity due to heating from SPAD voltage pulses. We expect material improvements to reduce dark counts and dark count variability between SPADs. Microfabricated optical filters could be used to reject light from sources at other wavelengths. An order of magnitude increase in collection efficiency may be achieved with larger active area and shielding that allows shuttling an ion directly over the detector. If some of these improvements are made (5\% collection efficiency, 100 cps DCR, elimination of background laser scattering) and the quantum efficiency remained the same (24\%), an average detection fidelity of 0.9977 could be achieved in 75~$\mu$s for state readout of the hyperfine S states typically used for quantum computing with $^{171}$Yb$^+$. We believe these trap-integrated detection results are a significant step towards enabling scalable and transportable quantum technologies.
\section{Acknowledgments}
\par The authors thank J. Hunker, B. Thurston, R. Haltli, J. Burkart, K. Setzer, and the members of Sandia’s Microsystems and Engineering Sciences Application (MESA) facility for their fabrication expertise and for helpful comments on the manuscript. This work is supported by the Defense Advanced Research Projects Activity (DARPA). Sandia National Laboratories is a multi-mission laboratory managed and operated by National Technology \& Engineering Solutions of Sandia, LLC, a wholly owned subsidiary of Honeywell International Inc., for the U.S. Department of Energy’s National Nuclear Security Administration under contract DE-NA0003525. This paper describes objective technical results and analysis. Any subjective views or opinions that might be expressed in the paper do not necessarily represent the views of the U.S. Department of Energy or the United States Government.
\section{Data Availability}
The data that support the findings of this study are available from the corresponding author upon reasonable request.
\bibliography{aip}
\end{document}